\documentclass[conference,a4paper]{IEEEtran}
\IEEEoverridecommandlockouts
\ifCLASSINFOpdf
\else
\fi
\usepackage{amsfonts,amssymb}
\usepackage{graphicx}
\usepackage{cite}
\usepackage{amssymb,amsfonts}
\usepackage{setspace}
\usepackage{textcomp}
\usepackage{color,soul}
\usepackage{multirow}
\usepackage{cuted}
\usepackage{array}
\usepackage{epsfig}
\usepackage{graphics}
\usepackage{varwidth}
\usepackage{amsmath}
\usepackage{breqn}
\usepackage{caption}
\usepackage{subcaption}
\usepackage{breqn}
\usepackage[T1]{fontenc}
\usepackage{mathtools}  
\usepackage{amsthm}
\usepackage{mathrsfs}
\usepackage{algorithm2e}
\usepackage{cleveref}
\allowdisplaybreaks
\usepackage{authblk}

\begin{document}

\title{Simultaneously Transmitting and Reflecting RIS-Aided Mobile Edge Computing: Computation Rate Maximization
}
\author[1,2]{\normalsize Zhenrong Liu}
\author[3]{Zongze Li}
\author[4]{Miaowen Wen}
\author[1]{Yi Gong}
\author[2]{Yik-Chung~Wu}
\affil[1]{Department of Electrical and Electronic Engineering, Southern University of Science and Technology, Shenzhen, China}
\affil[2]{Department of Electrical and Electronic Engineering, The University of Hong Kong, Hong Kong}
\affil[3]{Peng Cheng Laboratory, Shenzhen, China}
\affil[4]{School of Electronic and Information Engineering, South China University of Technology, Guangzhou, China}
\affil[ ]{Email: \textit {12050022@mail.sustech.edu.cn; lizz@pcl.ac.cn; eemwwen@scut.edu.cn; gongy@sustech.edu.cn; ycwu@eee.hku.hk}}
\maketitle

\begin{abstract}
In this paper, the novel simultaneously transmitting and reflecting (STAR) reconfigurable intelligent surface (RIS), which enables full-space coverage on users located on both sides of the surface, is investigated in the multi-user mobile edge computing (MEC) system. A computation rate maximization problem is formulated via the joint design of the STAR-RIS phase shifts, reflection and transmission amplitude coefficients, the receive beamforming vectors at the access point, and the users' energy partition strategies for local computing and offloading. Two operating protocols of STAR-RIS, namely energy splitting (ES) and mode switching (MS) are studied. Based on DC programming and semidefinite relaxation, an iterative algorithm is proposed for the ES protocol to solve the formulated non-convex problem. Furthermore, the proposed algorithm is extended to solve the non-convex, non-continuous MS problems with binary amplitude coefficients. Simulation results show that the resultant STAR-RIS-aided MEC system significantly improves the computation rate compared to the baseline scheme with conventional reflect-only/transmit-only RIS.
\end{abstract}

\section{Introduction}
Due to a plethora of emerging computation-intensive applications, there has been an unprecedented increase in computing demands on massive devices. In order to free resource-constrained devices from heavy computing workloads and provide them with high-performance, low-latency computing services, mobile edge computing (MEC) facilitates the use of cloud computing at the edge of mobile networks by integrating MEC servers at the access point (AP)\cite{7123563,9770199}.

Since the edge server has a powerful computation ability and the user's computing results are in small size, the bottleneck of the wireless MEC system is the uplink offloading performance\cite{8264794,7762913,8334188}. Unfortunately, the uplink offloading performance is severely limited in practice due to the energy constraints of the mobile users and adverse wireless channel conditions. To improve the performance of uplink offloading in MEC, reconfigurable intelligent surface (RIS) technology has attracted much attention due to its low cost, easy deployment, and propagation path improvement  \cite{8910627,9360709}. However, conventional RIS can only reflect the incident signal, which means that the AP and end users must be on the same side of the RIS, severely limiting the flexibility of the deployment and effectiveness of the RIS. A new concept of simultaneous transmission and reflection RIS (STAR-RIS) has recently emerged to overcome this geographical restriction\cite{9690478,9437234}. Inspired by the benefits of STAR-RIS, we integrate it into wireless MEC to break the geographical limitations of traditional RIS to provide full space coverage with better MEC services for end users.

In this paper, we propose using STAR-RIS to improve the performance of a wireless-aided MEC system. In particular, we formulate the computation rate maximization problem by considering both the local and offloaded computation rate with the user's available energy budget. However, the resultant problem is highly-nonconvex and non-continuous due to the coupling of the STAR-RIS's parameters with communication and computing resources and the binary reflection and transmission amplitude coefficients of STAR-RIS in mode-switching protocols. To address the above challenges, we transform the binary constraints into their equivalent continuous form and subsequently enforce their discrete binary values. Then, the resultant problem can be decomposed into three subproblems and iteratively solved under the block coordinate descent (BCD) framework, where semidefinite relaxation (SDR) and DC programming are used to handle each subproblem. Simulation results show that the proposed STAR-RIS-aided MEC system outperforms systems with the conventional RIS.

\section{System Model and Problem Formulation}
\label{sym}

\subsection{System Model}
\begin{figure}[t]
\centering
\centerline{\includegraphics[scale=0.07]{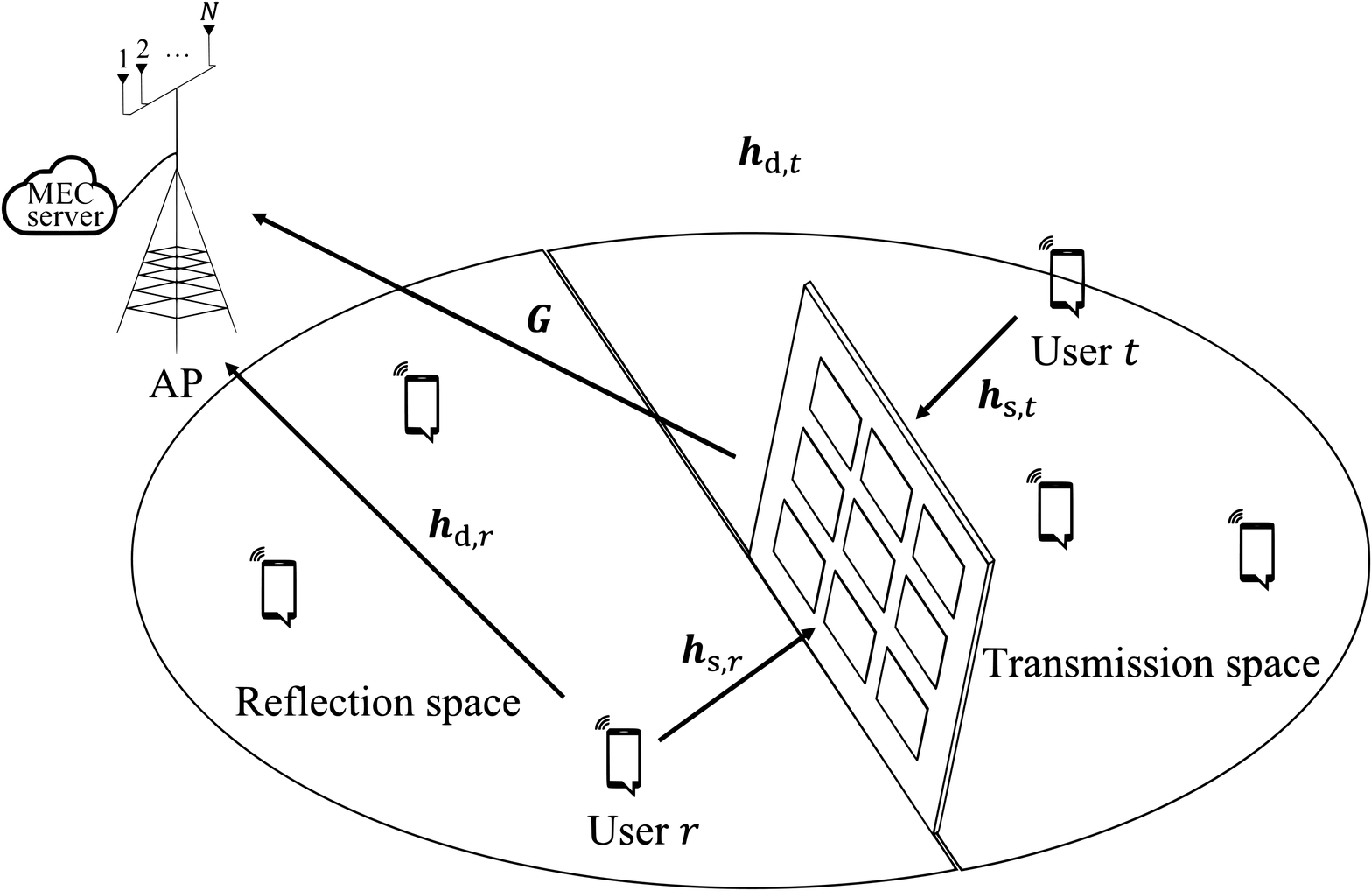}}
\caption{The RIS-aided MEC system.}
\label{sm}
\end{figure}

We consider a STAR-RIS-aided MEC system as shown in Fig. \ref{sm}, in which there are an $N$ antenna AP, $K$ single-antenna users, and an $M$-element STAR-RIS. The AP is attached to a MEC edge server. Since the STAR-RIS can provide full-space coverage by allowing simultaneous transmission and reflection of the incident signal, it serves both T users in the transmission space and R users in the reflection space. 

Each user has a limited energy budget $E_k$ but has intensive computation tasks. Therefore, we adopt the partial offloading mode to handle users' computation tasks with parallel local computing and computation offloading \cite{8016573}. For computation offloading, let $\mathcal{T}\!=\!\{1, \ldots, T\}$, $\mathcal{R}=$ $\{1, \ldots, R\}$ and $\mathcal{K}=\{\mathcal{T}\cup \mathcal{R}\}$ denote the index sets of T users, R users and all users, respectively. Let $s_{t}$ and $s_{r} \in \mathbb{C}$ with zero mean and unit variance denote the information symbol of T users $t \in \mathcal{T}$ and R users $ r \in \mathcal{R}$ for the offloading task, and $p_{t}$ and $p_{r}$ denote the transmit power of T users and R users, respectively. Note that all the users with offloading requirements are allowed to communicate with the edge server simultaneously, and thus we can express the corresponding received signal $\boldsymbol{y} \in \mathbb{C}^{L\times 1}$ at the AP as
\begin{equation}
\boldsymbol{y}=\sum_{t \in \mathcal{T}} \boldsymbol{g}_{t} \sqrt{p_{t}} s_{t}+\sum_{r \in \mathcal{R}} \boldsymbol{g}_{r} \sqrt{p_{r}} s_{r}+\boldsymbol{z},
\end{equation}
where $\boldsymbol{g}_{t}$ and $\boldsymbol{g}_{r}$ $\in \mathbb{C}^{L \times 1}$ are the equivalent baseband channel from user $t$ and $r$ to AP and $\boldsymbol{z} \sim \mathcal{C N}\left(\mathbf{0}, \sigma^{2} \boldsymbol{I}_{N}\right)$ is
the receiver noise at AP with $\sigma^{2}$ being the noise power. With the deployment of a STAR-RIS, the equivalent baseband channel from user $t$ or $r$ to the AP consists of both the direct and transmitting or reflecting links. Therefore, $\boldsymbol{g}_{t}$ can be modeled as
\begin{equation}
\label{T-channel}
\boldsymbol{g}_{t}=\boldsymbol{h}_{\mathrm{d}, t}+\left(\boldsymbol{G}\right)^{\mathrm{H}}\boldsymbol{\Theta}_{\mathrm{T}} \boldsymbol{h}_{\mathrm{s}, t},  \quad \forall t \in \mathcal{T},
\end{equation}
where $\boldsymbol{h}_{\mathrm{d},t} \in \mathbb{C}^{L \times 1}, \boldsymbol{h}_{\mathrm{s}, t} \in \mathbb{C}^{M \times 1}$, and $\boldsymbol{G} \in \mathbb{C}^{M \times L}$ denote the narrow-band quasi-static fading channels from user $t$ to AP, from user $t$ to the STAR-RIS, and from the STAR-RIS to AP, respectively. The transmission RIS matrix $\boldsymbol{\Theta}_{\mathrm{T}}=\operatorname{diag}\left(\rho^{\mathrm{t}}_1e^{j\theta_{1}}, \ldots, \rho^{\mathrm{t}}_Me^{j\theta_{M}}\right) \in \mathbb{C}^{M \times M}$ is a diagonal matrix, where $\rho^{\mathrm{t}}_m$ is the transmission amplitude coefficient and  $\theta_{m} \in[0,2 \pi)$ is the phase shift of the $m^{th}$ element’s transmission and reflection
coefficients.

Similar to (\ref{T-channel}), $\boldsymbol{g}_{r}$ is expressed as
\begin{equation}
\label{R-channel}
\boldsymbol{g}_{r}=\boldsymbol{h}_{\mathrm{d}, r}+\left(\boldsymbol{G}\right)^{\mathrm{H}}\boldsymbol{\Theta}_{\mathrm{R}} \boldsymbol{h}_{\mathrm{s}, r},  \quad \forall r \in \mathcal{R},
\end{equation}
where $\boldsymbol{h}_{\mathrm{d},r} \in \mathbb{C}^{L \times 1}$ and $\boldsymbol{h}_{\mathrm{s}, r} \in \mathbb{C}^{M \times 1}$ denote the equivalent baseband channel from user $r$ to AP, and from user $r$ to the STAR-RIS, respectively. The transmission RIS matrix $\boldsymbol{\Theta}_{\mathrm{R}}=\operatorname{diag}\left(\rho^{\mathrm{r}}_1e^{j\theta_{1}}, \ldots, \rho^{\mathrm{r}}_Me^{j\theta_{M}}\right) \in \mathbb{C}^{M \times M}$ is a diagonal matrix, where $\rho^{\mathrm{r}}_m$ is the reflection amplitude coefficient. 
This paper considers two protocols for operating the STAR-RIS, i.e., energy splitting (ES) and mode switching (MS). To be specific, for the ES mode, $\rho_{m}^{\mathrm{t}}, \rho_{m}^{\mathrm{r}} \in[0,1]$ and $\rho_{m}^{\mathrm{t}}+\rho_{m}^{\mathrm{r}}=1$. In terms of the MS mode, it restricts the reflection amplitude and transmission amplitude to be binary variables, i.e., $\rho_{m}^{\mathrm{t}}, \rho_{m}^{\mathrm{r}} \in \{0,1\}$. 





Next, we introduce an energy partition parameter $a_{k} \in[0,1]$ for users $k \in \mathcal{K}$. In computation offloading, $a_{k} E_{k}$ represents the energy allocated for computation offloading. In this case, the transmit power of user $k$ for computation offloading is given as $p_{k}={a_{k} E_{k}}/{L}\triangleq a_{k}\widetilde{E}_{k}$, where $\widetilde{E}_{k}=E_{k} / L$ and $L$ is the length of the time slot.

We consider the linear beamforming strategy and denote $\boldsymbol{v}_{k} \in$ $\mathbb{C}^{L \times 1}$ as the receive beamforming vector of AP for decoding $s_{k}$. The received signal at AP for user $k$, denoted by ${\hat{s}}_{k} \in \mathbb{C}$, is then given by
\begin{equation}
{\hat{s}}_{k}=\left(\boldsymbol{v}_{k}\right)^{\mathrm{H}} \boldsymbol{g}_{k} \sqrt{p_{k}} s_{k}+\left(\boldsymbol{v}_{k}\right)^{\mathrm{H}} \sum_{l \neq k} \boldsymbol{g}_{l} \sqrt{p_{l}}s_l+\left(\boldsymbol{v}_{k}\right)^{\mathrm{H}} \boldsymbol{z}.
\end{equation}
The uplink signal-to-interference-plus-noise ratio observed at AP for user $k$ is thus given by
\begin{equation}
\label{upsinr}
\!\!\!\gamma_{k}\Big(\!\{a_k\}, \boldsymbol{v}_{k},\boldsymbol{\Theta}_{\mathrm{T}},\boldsymbol{\Theta}_{\mathrm{R}}\!\Big)\!=\!\frac{p_{k}\left|\left(\boldsymbol{v}_{k}\right)^{\mathrm{H}}\!
\boldsymbol{g}_{k}
\right|^{2}}{\sum_{l \neq k} \!p_{l}\left|\left(\boldsymbol{v}_{k}\right)^{\mathrm{H}} \boldsymbol{g}_{l}
\right|^{2}\!\!\!+\!\sigma_{n}^{2}\left\|\boldsymbol{v}_{k}\right\|^{2}}.
\end{equation}
Then the computation rate of user $k$ offloaded to AP is
\begin{equation}
R_{k}\Big(\{a_k\}, \boldsymbol{v}_{k},\boldsymbol{\Theta}_{\mathrm{T}},\boldsymbol{\Theta}_{\mathrm{R}}\Big)=B\log _{2}\Big(1+\gamma_{k}\Big),\quad \forall k \in \mathcal{K},
\end{equation}
where $B$ is the bandwidth of the system.

As for the case of local computing, the dynamic voltage and frequency scaling technique is adopted for users to increase the computation energy efficiency through adaptively controlling the CPU frequency \cite{6574874}.
In particular, the computation energy consumption of user $k \in \mathcal{K}$ can be expressed as $V\kappa_kf_k^3$, where $f_k$ is the CPU frequency, and $\kappa_k$ is the effective capacitance coefficient of user $k$, respectively. As $(1-a_k)E_k$ is the energy for local computing, we have $(1-a_k)E_k\!=\!V\kappa_kf_k^3$, and thus we can calculate $f_k$ as $f_{k}\!=\!\sqrt[3]{\left((1-a_{k})E_{k}\right)/V\kappa_k}$. Let $C_k$ be the amount of required computing resource (i.e., the number of CPU cycles) for computing 1-bit of the user’s input data; the computation rate of user $k$ for local computing is
\begin{equation}
R_{k}^{\mathrm{loc}}\left(a_{k}\right)=\frac{f_{k} }{C_{k}}=\frac{1}{C_{k}} \sqrt[3]{\frac{(1-a_{k})E_{k}}{V\kappa_k}}, \quad \forall k \in \mathcal{K}.
\end{equation}

\subsection{Problem Formulation}
Our object is to maximize the computation rate of the users with energy budget $\left\{E_{k}\right\}_{k \in \mathcal{K}}$, including via computation offloading and local computing. This leads to the optimization problem being formulated as
\begin{subequations}
\label{originorigin}
\begin{align}
\label{origin-obj}
\!\!\mathscr{P}_{0}\!:\!\max _{\substack{\{a_k\}, \{\boldsymbol{v}_{k}\},\\ \boldsymbol{\Theta}_{\mathrm{T}},\boldsymbol{\Theta}_{\mathrm{R}}}}
&\!\sum_{k=1}^{K}\!\biggl(\!\!R_{k}\!\Big(\!\{a_k\}, \!\boldsymbol{v}_{k},\!\boldsymbol{\Theta}_{\mathrm{T}},\!\boldsymbol{\Theta}_{\mathrm{R}}\Big)+R_{k}^{\mathrm{loc}}\left(a_{k}\right)\biggr),\\
\text { s.t. } \quad \quad & a_{k} \in[0,1], \quad \forall k \in \mathcal{K}, \\
\label{MS}
& \rho_{m}^{\mathrm{t}}, \rho_{m}^{\mathrm{r}} \in\{0,1\}, \  \forall m \in \mathcal{M}, \text{(MS mode)}\\
\label{ES}
& \rho_{m}^{\mathrm{t}}, \rho_{m}^{\mathrm{r}} \in[0,1], \  \forall m \in \mathcal{M}, \text{(ES mode)}\\
&\rho_{m}^{\mathrm{t}}+\rho_{m}^{\mathrm{r}}=1, \quad \forall m \in \mathcal{M},\\
\label{mc1}
&\left|(\boldsymbol{\Theta}_{\mathrm{T}})_{m,m}\right|=\rho_{m}^{\mathrm{t}},\quad \forall m \in \mathcal{M},\\
\label{mc2}
&\left|(\boldsymbol{\Theta}_{\mathrm{R}})_{m,m}\right|=\rho_{m}^{\mathrm{r}},\quad \forall m \in \mathcal{M}.
\end{align}
\end{subequations}
Note that the constraints (\ref{ES}) and (\ref{MS}) are mutually exclusive for $\mathscr{P}_{0}$ as they represent different operation protocols for STAR-RIS. To be specific, the STAR-RIS in MS mode should be constrained by (\ref{MS}) rather than (\ref{ES}) and vice verses.  

Due to the coupling of optimization variables and the modulus constraints (\ref{mc1}), (\ref{mc2}), it is difficult to solve $\mathscr{P}_{0}$ in general. Furthermore, since the amplitude coefficients $\rho_{m}^{\mathrm{t}}$ and $\rho_{m}^{\mathrm{r}}$ are binary variables in MS mode, the objective function (\ref{origin-obj}) is discontinuous and it usually requires the exponential time complexity to find the optimal solution. To address these issues, we first optimize $\mathscr{P}_{0}$ under the BCD framework to effectively separate the coupling among the optimization variables, thereby facilitating the subsequent algorithm design.

\section{Block Coordinate Descent Algorithm Design}
\label{3}
Under the BCD framework, the details of optimization algorithms for solving each subproblem are derived. 
\subsection{Solving $\{\boldsymbol{v}_{k}\}$}
When other variables are fixed, the subproblem of $\mathscr{P}_{0}$ for updating $\{\boldsymbol{v}_{k}\}$ is
\begin{subequations}
\label{vk}
\begin{align}
\max _{\substack{\{\boldsymbol{v}_{k}\}}} \quad
&\sum_{k=1}^{K}R_{k}\left(\boldsymbol{v}_{k}\right).
\end{align}
\end{subequations}
Since the $k^{th}$ users computation rate for offloading only contains its receive beamforming vector $\boldsymbol{v}_k$, we can therefore optimize it for each $k$ separately, with $k^{th}$ subproblem as
$
\max _{\boldsymbol{v}_{k}} \gamma_{k}\left(\boldsymbol{v}_{k}\right)=\frac{\boldsymbol{v}_{k}^{\mathrm{H}} \boldsymbol{A}_{k} \boldsymbol{v}_{k}}{\boldsymbol{v}_{k}^{\mathrm{H}} \boldsymbol{B}_{k} \boldsymbol{v}_{k}},
$
where $\boldsymbol{A}_{k}=p_{k} \boldsymbol{g}_{k}\left(\boldsymbol{g}_{k}\right)^{\mathrm{H}}$ and $\boldsymbol{B}_{k}=\sum_{i=1, i \neq n}^{K} p_{i} \boldsymbol{g}_{i}\left(\boldsymbol{g}_{i}\right)^{\mathrm{H}}+$ $\sigma^{2} \mathbf{I}_{N}$. By observing that it is a generalized eigenvector problem, the optimal solution $\boldsymbol{v}_{k}^{*}$ should be the eigenvector corresponds to the largest eigenvalue of the matrix $\left(\boldsymbol{B}_{k}\right)^{-1} \boldsymbol{A}_{k}$\cite{9380744}. 

\subsection{Optimizing STAR-RIS Matrix $\boldsymbol{\Theta}_{\mathrm{T}},\boldsymbol{\Theta}_{\mathrm{R}}$}

When it comes to the STAR-RIS matrix design, the subproblem for updating $\boldsymbol{\Theta}_{\mathrm{T}},\boldsymbol{\Theta}_{\mathrm{R}}$ is given by
\begin{subequations}
\label{thetatr}
\begin{align}
\label{obj-theta}
\max _{\substack{ \boldsymbol{\Theta}_{\mathrm{T}}}, \boldsymbol{\Theta}_{\mathrm{R}}} \quad
&\sum_{t=1}^{T}R_{t}\left(\boldsymbol{\Theta}_{\mathrm{T}},\boldsymbol{\Theta}_{\mathrm{R}}\right)+\sum_{r=1}^{R}R_{r}\left(\boldsymbol{\Theta}_{\mathrm{T}},\boldsymbol{\Theta}_{\mathrm{R}}\right),\\
\text { s.t. } \quad 
& \rho_{m}^{\mathrm{t}}, \rho_{m}^{\mathrm{r}} \in\{0,1\}, \  \forall m \in \mathcal{M}, \text{(MS mode)}\\
& \rho_{m}^{\mathrm{t}}, \rho_{m}^{\mathrm{r}} \in[0,1], \  \forall m \in \mathcal{M}, \text{(ES mode)}\\
&\rho_{m}^{\mathrm{t}}+\rho_{m}^{\mathrm{r}}=1,\quad \forall m \in \mathcal{M},\\
&\left|(\boldsymbol{\Theta}_{\mathrm{T}})_{m,m}\right|=\rho_{m}^{\mathrm{t}},\quad \forall m \in \mathcal{M},\\
&\left|(\boldsymbol{\Theta}_{\mathrm{R}})_{m,m}\right|=\rho_{m}^{\mathrm{r}},\quad \forall m \in \mathcal{M}.
\end{align}
\end{subequations}

According to the expression of $g_t,g_r$ in (\ref{T-channel}) and (\ref{R-channel}), we can re-express $\left|\left(\boldsymbol{v}_{k}\right)^{\mathrm{H}} \boldsymbol{g}_{k}\right|^{2}$ as
\begin{equation}
\begin{aligned}
\left|\left(\boldsymbol{v}_{k}\right)^{\mathrm{H}}\!\!\boldsymbol{g}_{k}\right|^{2}\! \!=&\left|h_{\mathrm{d},k,k}+\left(\boldsymbol{v}_{k}\right)^{\mathrm{H}}(\boldsymbol{G})^{\mathrm{H}}\operatorname{diag}(\boldsymbol{h}_{\mathrm{s}, k}) \boldsymbol{\phi}_{\mathrm{T}/\mathrm{R}}\right|^{2},\\
=&\left|h_{\mathrm{d},k,k}+\boldsymbol{h}_{\mathrm{s},k,k} \boldsymbol{\phi}_{\mathrm{T}/\mathrm{R}}\right|^{2},
\end{aligned}
\end{equation}
where $\!h_{\mathrm{d},k,k}\!=\!\left(\boldsymbol{v}_{k}\right)^{\mathrm{H}}\boldsymbol{h}_{\mathrm{d}, k}$, $\boldsymbol{h}_{\mathrm{r},k,k}\!=\!\left(\boldsymbol{v}_{k}\right)^{\mathrm{H}}\!(\boldsymbol{G})^{\mathrm{H}}\operatorname{diag}(\boldsymbol{h}_{\mathrm{r}, k}) \in \mathbb{C}^{1 \times M}$ and $\boldsymbol{\phi}_{\mathrm{T}/\mathrm{R}}=\Big[\rho^{\mathrm{t}/\mathrm{r}}_1e^{j\theta_{1}^{\mathrm{t}/\mathrm{r}}}, \ldots, \allowbreak \rho^{\mathrm{t}/\mathrm{r}}_Me^{j\theta_{M}^{\mathrm{t}/\mathrm{r}}}\Big]^{\mathrm{T}} \in \mathbb{C}^{M \times 1}$. By defining a matrix $\boldsymbol{Q}_{k,k} \in \mathbb{C}^{(M+1) \times(M+1)}$ as
\begin{equation}
\boldsymbol{Q}_{k,k}=\left[\begin{array}{lc}
\left(\boldsymbol{h}_{\mathrm{s},k,k}\right)^{\mathrm{H}} \boldsymbol{h}_{\mathrm{s},k,k} & \left(\boldsymbol{h}_{\mathrm{s},k,k}\right)^{\mathrm{H}} h_{\mathrm{d},k,k} \\
h_{\mathrm{d},k,k}^{\mathrm{H}} \boldsymbol{h}_{\mathrm{s},k,k} & 0
\end{array}\right],
\end{equation}
and a vector $\widetilde{\boldsymbol{\phi}}_{\mathrm{T},\mathrm{R}}=\left[\boldsymbol{\phi}_{\mathrm{T}/\mathrm{R}}^{\mathrm{T}}, 1\right]^{\mathrm{T}} \in \mathbb{C}^{(K+1) \times 1}$, we can then re-express $\left|h_{\mathrm{d},k,k}+\boldsymbol{h}_{\mathrm{s},k,k} \boldsymbol{\phi}_{\mathrm{T}/\mathrm{R}}\right|^{2}=$ $\tilde{\boldsymbol{\phi}}_{\mathrm{T},\mathrm{R}}^{\mathrm{H}} \boldsymbol{Q}_{k,k} \tilde{\boldsymbol{\phi}}_{\mathrm{T},\mathrm{R}}+\left|h_{\mathrm{d},k, k}\right|^{2}=\operatorname{Tr}\left(\boldsymbol{Q}_{k,k} \boldsymbol{\Psi}_{\mathrm{T}/\mathrm{R}}\right)+\left|h_{\mathrm{d}, k, k}\right|^{2}$, where $\boldsymbol{\Psi}_{\mathrm{T}/\mathrm{R}}=$ $\widetilde{\boldsymbol{\phi}}_{\mathrm{T},\mathrm{R}} \widetilde{\boldsymbol{\phi}}_{\mathrm{T},\mathrm{R}}^{\mathrm{H}} \in \mathbb{C}^{(K+1) \times(K+1)}$ is a positive semidefinite matrix related to the STAR-RIS reflecting coefficients. Therefore each added item in the (\ref{obj-theta}), i.e., $R_{k}\left(\boldsymbol{\Theta}_{\mathrm{T}/\mathrm{R}}\right)$, can be re-written as
\begin{equation}
\begin{aligned}
&\!\!\log _{2}\left(1+\gamma_{k}\left(\boldsymbol{\Theta}_{\mathrm{T}/\mathrm{R}}\right)\right)=\log _{2}\left(1+\gamma_{k}\left(\boldsymbol{\Psi}_{\mathrm{T}/\mathrm{R}}\right)\right) \\
=&\log_{2}\!\!\left(\!\sum_{j=1}^{K} \!p_{j}\!\left(\!\operatorname{Tr}\left(\boldsymbol{Q}_{k,j} \boldsymbol{\Psi}_{\mathrm{T}/\mathrm{R}}\right)\!\!+\!\!\left|h_{\mathrm{d}, k, j}\right|^{2}\!\right)\!\!+\!\!\sigma^{2}\!\!\left\|\boldsymbol{v}_{k}\right\|^{2}\!\!\right) \\
-&\!\log_{2}\!\!\left(\!\sum_{i=1, i \neq k}^{K} \!\!\!\!\!p_{i}\!\!\left(\!\operatorname{Tr}\!\left(\boldsymbol{Q}_{k, i} \boldsymbol{\Psi}_{\mathrm{T}/\mathrm{R}}\right)\!\!+\!\!\left|h_{\mathrm{d}, k, i}\right|^{2}\!\right)\!\!+\!\!\sigma^{2}\left\|\boldsymbol{v}_{k}\right\|^{2}\!\!\right) \\
\triangleq & F_{1, k}(\boldsymbol{\Psi}_{\mathrm{T}/\mathrm{R}})-F_{2, k}(\boldsymbol{\Psi}_{\mathrm{T}/\mathrm{R}}), \quad \forall k \in \mathcal{K},
\end{aligned}
\end{equation}
where $F_{1, k}(\boldsymbol{\Psi}_{\mathrm{T}/\mathrm{R}})$ and $F_{2, k}(\boldsymbol{\Psi}_{\mathrm{T}/\mathrm{R}})$ are two concave functions with respect to $\boldsymbol{\Psi}_{\mathrm{T}/\mathrm{R}}$. Hence, (\ref{thetatr}) can be equivalently transformed
into the following problem\cite{2010.15354}
\begin{subequations}
\label{DC-theta}
\begin{align}
\label{DC-obj}
\max _{\substack{\boldsymbol{\Psi}_{\mathrm{T}/\mathrm{R}} \succeq \mathbf{0},\\\boldsymbol{\rho}^{\mathrm{T}/\mathrm{R}}}} \quad & \sum_{k=1}^{K} F_{1, k}(\boldsymbol{\Psi}_{\mathrm{T}/\mathrm{R}})-F_{2, k}(\boldsymbol{\Psi}_{\mathrm{T}/\mathrm{R}}) \\
\text { s.t. } \quad & \operatorname{diag}\left({\boldsymbol{\Psi}_{\mathrm{T}/\mathrm{R}}}\right)=\boldsymbol{\rho}^{\mathrm{T}/\mathrm{R}},\\
\label{rank}
& \operatorname{rank}\left({\boldsymbol{\Psi}_{\mathrm{T}/\mathrm{R}}}\right)=1,\\
\label{ms1}
& \rho_{m}^{\mathrm{t}}, \rho_{m}^{\mathrm{r}} \in\{0,1\}, \  \forall m \in \mathcal{M}, \text{(MS mode)}\\
\label{es1}
& \rho_{m}^{\mathrm{t}}, \rho_{m}^{\mathrm{r}} \in[0,1], \  \forall m \in \mathcal{M}, \text{(ES mode)}\\
&\rho_{m}^{\mathrm{t}}+\rho_{m}^{\mathrm{r}}=1,\quad \forall m \in \mathcal{M},
\end{align}
\end{subequations}
where $\boldsymbol{\rho}^{\mathrm{T}/\mathrm{R}} \triangleq$ $\left[\rho_{1}^{\mathrm{t}/\mathrm{r}}, \rho_{2}^{\mathrm{t}/\mathrm{r}}, \ldots, \rho_{M}^{\mathrm{t}/\mathrm{r}}\right]^{\mathrm{T}}$.

Note that the non-convexity of the problem (\ref{DC-theta}) comes from the objective function in (\ref{DC-obj}), the rank-one constraints (\ref{rank}), and the binary constraints for modeling MS mode (\ref{ms1}). Since the objective function is a sum of differences between two concave functions, we will show that the DC programming can be leveraged to effectively address the non-convexity of the objective function, rank-one constraints, and binary constraints\cite{an2005dc}.

As for the objective function, in the $(l+1)^{th}$ iteration of the DC programming, the second concave item, i.e., $F_{2, k}(\boldsymbol{\Psi}_{\mathrm{T}/\mathrm{R}})$ for $k \in \mathcal{K}$, can be locally approximated by its linear upper bound at the point $\boldsymbol{\Psi}_{\mathrm{T}/\mathrm{R}}^{(l)}$ (the solution obtained from the previous $l^{th}$ iteration), which is given as
\begin{equation}
\label{DC-objdc}
\begin{aligned}
&F_{2, k}(\boldsymbol{\Psi}_{\mathrm{T}/\mathrm{R}}) \leq \widehat{F}_{2, k}\left(\boldsymbol{\Psi}_{\mathrm{T}/\mathrm{R}}; \boldsymbol{\Psi}_{\mathrm{T}/\mathrm{R}}^{(l)}\right)=F_{2, k}\left(\boldsymbol{\Psi}_{\mathrm{T}/\mathrm{R}}^{(l)}\right)+ \\
&\frac{\sum\limits_{i=1, i \neq k}^{K} \!\!\!\!\!p_{i}\!\!\left\langle\!\!\!\left(\!\boldsymbol{\Psi}_{\mathrm{T}/\mathrm{R}}\!\!-\!\! \boldsymbol{\Psi}_{\mathrm{T}/\mathrm{R}}^{(l)}\right)\!,\!\!\left.\nabla_{\boldsymbol{\Psi}_{\mathrm{T}/\mathrm{R}}} \!\!\operatorname{Tr}\!\left(\mathbf{Q}_{k, i} \boldsymbol{\Psi}_{\mathrm{T}/\mathrm{R}}\right)\!\right|_{\boldsymbol{\Psi}_{\mathrm{T}/\mathrm{R}}=\boldsymbol{\Psi}_{\mathrm{T}/\mathrm{R}}^{(l)}}\!\right\rangle}{\ln 2\left(\sum\limits_{i=1, i \neq k}^{K} p_{i}\left(\operatorname{Tr}\left(\mathbf{Q}_{k, i} \boldsymbol{\Psi}_{\mathrm{T}/\mathrm{R}}^{(l)}\right)+\left|h_{\mathrm{d}, k, i}\right|^{2}\right)+\sigma^{2}\left\|\boldsymbol{v}_{k}\right\|^{2}\right)},
\end{aligned}
\end{equation}
where \!$\!\left.\nabla_{\!\boldsymbol{\Psi}_{\mathrm{T}/\mathrm{R}}}\!\operatorname{Tr}\!\left(\mathbf{Q}_{k, i} \boldsymbol{\Psi}_{\mathrm{T}/\mathrm{R}}\right)\right|_{\boldsymbol{\Psi}_{\mathrm{T}/\mathrm{R}}=\boldsymbol{\Psi}_{\mathrm{T}/\mathrm{R}}^{(l)}}\!\!$ denotes the Jacobian matrix of $\operatorname{Tr}\left(\mathbf{Q}_{k, i} \boldsymbol{\Psi}_{\mathrm{T}/\mathrm{R}}\right)$ with respect to $\boldsymbol{\Psi}_{\mathrm{T}/\mathrm{R}}$ at the point $\boldsymbol{\Psi}_{\mathrm{T}/\mathrm{R}}^{(l)}$, and it is easy to note that the equality holds when $\boldsymbol{\Psi}_{\mathrm{T}/\mathrm{R}}=\boldsymbol{\Psi}_{\mathrm{T}/\mathrm{R}}^{(l)}$.

As for the rank-one constraints, it can be equivalently transformed into the following form
\begin{equation}
\operatorname{Tr}(\boldsymbol{\Psi}_{\mathrm{T}/\mathrm{R}})-\|\boldsymbol{\Psi}_{\mathrm{T}/\mathrm{R}}\|_{\mathrm{s}}=0,
\end{equation}
where $\|\boldsymbol{\Psi}_{\mathrm{T}/\mathrm{R}}\|_{\mathrm{s}}$ denotes the spectral norm of the matrix $\boldsymbol{\Psi}_{\mathrm{T}/\mathrm{R}}$. It is noticeable that $\operatorname{Tr}(\boldsymbol{\Psi}_{\mathrm{T}/\mathrm{R}})=\sum_{m=1}^{M+1} \beta_{m}(\boldsymbol{\Psi}_{\mathrm{T}/\mathrm{R}})$ and $\|\boldsymbol{\Psi}_{\mathrm{T}/\mathrm{R}}\|_{\mathrm{s}}=$ $\beta_{1}(\boldsymbol{\Psi}_{\mathrm{T}/\mathrm{R}})$, where $\beta_{m}(\boldsymbol{\Psi}_{\mathrm{T}/\mathrm{R}})$ indicates the $m^{th}$ largest singular value of $\boldsymbol{\Psi}_{\mathrm{T}/\mathrm{R}}$. Hence, the equality of $\operatorname{Tr}(\boldsymbol{\Psi}_{\mathrm{T}/\mathrm{R}})=\|\boldsymbol{\Psi}_{\mathrm{T}/\mathrm{R}}\|_{\mathrm{s}}$ holds when the rank-one constraint is satisfied with $\beta_{1}(\boldsymbol{\Psi}_{\mathrm{T}/\mathrm{R}})>0$ and $\beta_{m}(\boldsymbol{\Psi}_{\mathrm{T}/\mathrm{R}})=$ 0 for $m=2, \ldots, M+1$, and vice versa. Similarly, in the $(l+1)^{th}$ iteration of the DC programming, a linear lower-bound of the convex item $\|\boldsymbol{\Psi}_{\mathrm{T}/\mathrm{R}}\|_{\mathrm{s}}$ at the point $\boldsymbol{\Psi}_{\mathrm{T}/\mathrm{R}}^{(l)}$ can be expressed as
\begin{equation}
\label{DC-rank1}
\begin{aligned}
&\|\boldsymbol{\Psi}_{\mathrm{T}/\mathrm{R}}\|_{\mathrm{s}} \geq\left\|\boldsymbol{\Psi}_{\mathrm{T}/\mathrm{R}}^{(l)}\right\|_{\mathrm{s}}\\
&+\!\left\langle\!\left(\!\boldsymbol{\Psi}_{\mathrm{T}/\mathrm{R}}-\boldsymbol{\Psi}_{\mathrm{T}/\mathrm{R}}^{(l)}\right),\left.\partial_{\boldsymbol{\Psi}_{\mathrm{T}/\mathrm{R}}}\|\boldsymbol{\Psi}_{\mathrm{T}/\mathrm{R}}\|_{\mathrm{s}}\right|_{\boldsymbol{\Psi}_{\mathrm{T}/\mathrm{R}}=\boldsymbol{\Psi}_{\mathrm{T}/\mathrm{R}}^{(l)}}\right\rangle \\
& \triangleq \Upsilon\left(\boldsymbol{\Psi}_{\mathrm{T}/\mathrm{R}}; \boldsymbol{\Psi}_{\mathrm{T}/\mathrm{R}}^{(l)}\right),
\end{aligned}
\end{equation}
where $\left.\partial_{\boldsymbol{\Psi}_{\mathrm{T}/\mathrm{R}}}\|\boldsymbol{\Psi}_{\mathrm{T}/\mathrm{R}}\|_{\mathrm{s}}\right|_{\boldsymbol{\Psi}_{\mathrm{T}/\mathrm{R}}=\boldsymbol{\Psi}_{\mathrm{T}/\mathrm{R}}^{(l)}}$ is a subgradient of the spectral norm $\|\boldsymbol{\Psi}_{\mathrm{T}/\mathrm{R}}\|$ with respect to $\boldsymbol{\Psi}_{\mathrm{T}/\mathrm{R}}$ at the point $\boldsymbol{\Psi}_{\mathrm{T}/\mathrm{R}}^{(l)}$, and the equality holds when $\boldsymbol{\Psi}_{\mathrm{T}/\mathrm{R}}=\boldsymbol{\Psi}_{\mathrm{T}/\mathrm{R}}^{(l)}$. Note that one subgradient of $\|\boldsymbol{\Psi}_{\mathrm{T}/\mathrm{R}}\|_{\mathrm{s}}$ at point $\boldsymbol{\Psi}_{\mathrm{T}/\mathrm{R}}^{(l)}$ can be efficiently computed as $\boldsymbol{z}_{1} \boldsymbol{z}_{1}^{\mathrm{H}}$, where $\boldsymbol{z}_{1}$ is the vector corresponding to the largest singular value of $\boldsymbol{\Psi}_{\mathrm{T}/\mathrm{R}}^{(l)}$\cite{8952884}.

With the obtained linear lower bound of $\|\boldsymbol{\Psi}_{\mathrm{T}/\mathrm{R}}\|_{\mathrm{s}}$ in (\ref{DC-rank1}), we can generate an approximate rank-one constraint of (\ref{rank}), which is shown as
\begin{equation}
\operatorname{Tr}(\boldsymbol{\Psi}_{\mathrm{T}/\mathrm{R}})-\Upsilon\left(\boldsymbol{\Psi}_{\mathrm{T}/\mathrm{R}}; \boldsymbol{\Psi}_{\mathrm{T}/\mathrm{R}}^{(l)}\right) \leq \varepsilon_{\Psi},
\end{equation}
where $\varepsilon_{\Psi}$ is a positive threshold with a very small value close to zero. The approximated rank-one constraint can guarantee that $0 \leq \operatorname{Tr}(\boldsymbol{\Psi}_{\mathrm{T}/\mathrm{R}})-\|\boldsymbol{\Psi}_{\mathrm{T}/\mathrm{R}}\|_{\mathrm{s}} \leq \operatorname{Tr}(\boldsymbol{\Psi}_{\mathrm{T}/\mathrm{R}})-\Upsilon\left(\boldsymbol{\Psi}_{\mathrm{T}/\mathrm{R}}; \boldsymbol{\Psi}_{\mathrm{T}/\mathrm{R}}^{(l)}\right) \leq \varepsilon_{\Psi}$, and the rank-one constraint can be approached with arbitrary accuracy by setting the value of $\varepsilon_{\Psi}$.

In terms of the non-convex binary constraints (\ref{ms1}) in MS mode, we first transform it equivalently into its continuous form as\cite{murray2010algorithm}
\begin{align}
\label{binary-qcqp}
&\rho_{m}^{\mathrm{x}}-\left(\rho_{m}^{\mathrm{x}}\right)^{2}=0, \quad \forall \mathrm{x} \in\{\mathrm{t}, \mathrm{r}\}, m \in \mathcal{M},\\
\label{0-1}
&0 \leq \rho_{m}^{\mathrm{t}}, \rho_{m}^{\mathrm{r}} \leq 1, \quad \forall \mathrm{x} \in\{\mathrm{t}, \mathrm{r}\}, m \in \mathcal{M}.
\end{align}
Due to the constraint (\ref{0-1}), we always have $\rho_{m}^{\mathrm{x}}-\left(\rho_{m}^{\mathrm{x}}\right)^{2}\geq0$, where equality holds if and only if $\rho_{m}^{\mathrm{x}}$ is $0$ or $1$, i.e., a binary variable. Similarly, in the $(l+1)^{th}$ iteration of the DC programming, a linear lower-bound of the convex item $\left(\rho_{m}^{\mathrm{x}}\right)^{2}$ at the point $\left(\rho_{m}^{\mathrm{x}}\right)^{(l)}$ can be expressed as
\begin{equation}
\label{DC-QCQP}
\begin{aligned}
\left(\rho_{m}^{\mathrm{x}}\right)^{2} &\geq\left(\left(\rho_{m}^{\mathrm{x}}\right)^{(l)}\right)^{2}+2 \left(\rho_{m}^{\mathrm{x}}\right)^{(l)}\left(\rho_{m}^{\mathrm{x}}-\left(\rho_{m}^{\mathrm{x}}\right)^{(l)}\right)\\
&=\Omega\left(\rho_{m}^{\mathrm{x}};\left(\rho_{m}^{\mathrm{x}}\right)^{(l)}\right),
\end{aligned}
\end{equation}
and the equality holds when $\rho_{m}^{\mathrm{x}}=\left(\rho_{m}^{\mathrm{x}}\right)^{(l)}$.

With the obtained linear lower bound of $\left(\rho_{m}^{\mathrm{x}}\right)^{2}$ in (\ref{DC-QCQP}), we can generate an approximated binary constraint of (\ref{binary-qcqp}), which is shown as
\begin{equation}
\rho_{m}^{\mathrm{x}}-\Omega\left(\rho_{m}^{\mathrm{x}};\left(\rho_{m}^{\mathrm{x}}\right)^{(l)}\right) \leq \varepsilon_{\rho},
\end{equation}
where $\varepsilon_{\rho}$ is a positive threshold with a very small value close to zero. The approximated binary constraint, together with the corresponding constraint in (\ref{0-1}) can guarantee that $0 \leq \rho_{m}^{\mathrm{x}}-\left(\rho_{m}^{\mathrm{x}}\right)^{2}\leq \rho_{m}^{\mathrm{x}}-\Omega\left(\rho_{m}^{\mathrm{x}};\left(\rho_{m}^{\mathrm{x}}\right)^{(l)}\right) \leq \varepsilon_{\rho}$, and the binary constraints can be approached with arbitrary accuracy by setting the value of $\varepsilon_{\rho}$.

To this end, we can obtain an approximation problem of (\ref{DC-theta}) at the $(l+1)^{th}$ iteration as
\begin{subequations}
\label{DC-theta-final}
\begin{align}
\max _{\substack{\boldsymbol{\Psi}_{\mathrm{T}/\mathrm{R}} \succeq \mathbf{0},\\\boldsymbol{\rho}^{\mathrm{T}/\mathrm{R}}}} \quad & \sum_{k=1}^{K} F_{1, k}(\boldsymbol{\Psi}_{\mathrm{T}/\mathrm{R}})-F_{2, k}(\boldsymbol{\Psi}_{\mathrm{T}/\mathrm{R}}) \\
\text { s.t. } \quad & \operatorname{diag}\left({\boldsymbol{\Psi}_{\mathrm{T}/\mathrm{R}}}\right)=\boldsymbol{\rho}^{\mathrm{T}/\mathrm{R}},\\
& \operatorname{Tr}(\boldsymbol{\Psi}_{\mathrm{T}/\mathrm{R}})-\Upsilon\left(\boldsymbol{\Psi}_{\mathrm{T}/\mathrm{R}}; \boldsymbol{\Psi}_{\mathrm{T}/\mathrm{R}}^{(l)}\right) \leq \varepsilon_{\Psi},\\
& \rho_{m}^{\mathrm{x}}-\Omega\left(\rho_{m}^{\mathrm{x}};\left(\rho_{m}^{\mathrm{x}}\right)^{(l)}\right) \leq \varepsilon_{\rho},\text{ (MS mode)}\\
&\rho_{m}^{\mathrm{t}}+\rho_{m}^{\mathrm{r}}=1,\quad \forall m \in \mathcal{M},\\
&0 \leq \rho_{m}^{\mathrm{t}}, \rho_{m}^{\mathrm{r}} \leq 1,\quad \forall m \in \mathcal{M},
\end{align}
\end{subequations}
which is a convex optimization problem and can be readily solved by the existing convex solvers such as CVX, and the optimal solution can be obtained as $\boldsymbol{\Psi}_{\mathrm{T}/\mathrm{R}}^{(l+1)}$.

\subsection{Optimizing Variables $\{a_k\}$}

When other variables are fixed, the subproblem for updating $\{a_k\}$ is
\begin{subequations}
\label{energyp}
\begin{align}
\label{energyp-obj}
\max _{\substack{\{a_k\}}} \quad
&\sum_{k=1}^{K}R_{k}\Big(\{a_k\}\Big)+\sum_{k=1}^{K}R_{k}^{\mathrm{loc}}\left(a_{k}\right),\\
\text { s.t. } \quad & a_{k} \in[0,1], \quad \forall k \in \mathcal{K}.
\end{align}
\end{subequations}
Note that problem (\ref{energyp}) is non-convex because of the non-concave items $R_{k}\left(\{a_k\}\right)$ in the objective function (\ref{energyp-obj}). Similarly, $R_{k}\left(\{a_k\}\right)$ for $k\in\mathcal{K}$ can be re-expressed as the difference of two concave functions as follows
\begin{equation}
\label{CCP}
\begin{aligned}
&R_{k}\Big(\{a_k\}\Big) \triangleq R_{k, 1}\Big(\{a_k\}\Big)-R_{k, 2}\Big(\{a_k\}\Big) \\
=& B \log _{2}\left(\sum_{j=1}^{K} a_{j} \widetilde{E}_{j}\left|\boldsymbol{v}_{k}^{\mathrm{H}} \boldsymbol{g}_{j}\right|^{2}+\sigma^{2}|| \boldsymbol{v}_{k} \|^{2}\right) \\
-&B \log _{2}\left(\sum_{i=1, i \neq k}^{K} a_{i} \widetilde{E}_{i}\left|\boldsymbol{v}_{k}^{\mathrm{H}} \boldsymbol{g}_{i}\right|^{2}+\sigma^{2}|| \boldsymbol{v}_{k} \|^{2}\right),
\end{aligned}
\end{equation}

Then the problem (\ref{energyp}) can also be solved by DC programming with the second term in (\ref{CCP}), i.e., $-R_{k, 2}\Big(\{a_k\}\Big)$, substituted by its upper bound to obtain a concave approximation of $R_{k}\left(\{a_k\}\right)$. Assuming $\{a_k^{(n)}\}$ is the solution obtained at the $n^{th}$ iteration of the DC programming, a linear upper bound of $R_{k, 2}\Big(\{a_k\}\Big)$ at the point $\{a_k^{(n)}\}$ can be obtained as
\begin{equation}
\begin{aligned}
&R_{k, 2}\Big(\{a_k\}\Big) \leq \widehat{R}_{k, 2}\left(\{a_k\}; \{a_k^{(n)}\}\right) \\
=& R_{k, 2}\Big(\{a_k^{(n)}\}\Big)+\!\!\!\!\sum_{i=1, i \neq k}^{K} R_{k, 2, i}^{\prime}\left(\{a_k^{(n)}\}\right)\left(a_{i}-a_{i}^{(n)}\right),
\end{aligned}
\end{equation}
where $R_{k, 2, i}^{\prime}\left(\{a_k^{(n)}\}\right) = \frac{B}{\ln 2} \frac{\widetilde{E}_{i}\left|\boldsymbol{v}_{k}^{\mathrm{H}} \boldsymbol{g}_{i}\right|^{2}}{\sum_{j=1, j \neq k}^{K} a_{j}^{(n)} \widetilde{E}_{j}\left|\boldsymbol{v}_{k}^{\mathrm{H}} \boldsymbol{g}_{j}\right|^{2}+\sigma^{2}\left\|\boldsymbol{v}_{k}\right\|^{2}}$ is the derivative of $R_{k, 2, i}^{\prime}\left(\{a_k^{(n)}\}\right)$ with respect to $a_i$ at the point $\{a_k^{(n)}\}$. It is easy to note that the equality holds when $\{a_k\}=\{a_k^{(n)}\}$. At the $(n+1)^{th}$ iteration of DC programming, we aim to maximize the following approximation problem
\begin{equation}
\label{energyp-final}
\begin{aligned}
\!\!\!\!\!\!\!\max_{\substack{\{a_k\}}} \quad &\!\!\!\!\!\sum_{k=1}^{K}\!\left(\!R_{k, 1}\Big(\!\{a_k\}\!\Big)\!\!-\!\!\widehat{R}_{k, 2}\!\!\left(\!\{a_k\}; \{a_k^{(n)}\}\!\right)\!\!+\!\!R_{k}^{\mathrm{loc}}\left(a_{k}\right)\!\right),\\
\text { s.t. } \quad &a_{k} \in[0,1], \quad \forall k \in \mathcal{K} \text {, }
\end{aligned}
\end{equation}
which is a convex problem and can be easily solved by CVX. Through solving problem (\ref{energyp-final}) with CVX, the optimal solution, i.e., $\{a_k^{(n+1)}\}$, can be finally obtained.

\section{Simulation Results}
In this section, we present simulation results to verify the effectiveness of our proposed algorithm. Under a three-dimensional Euclidean coordinate system, we consider a system with four users in transmission space and four users in reflection space, and they are randomly located in a square region of $50 \mathrm{~m} \times 50 \mathrm{~m}$ centered at the 3-dimensional coordinate $(45,0,0)$ and $(95,0,0)$, respectively. A STAR-RIS and an AP are located at the 3-dimensional coordinate $(75,0,15)$ and $(0,0,15)$, respectively.

Rician fading channel is considered to model both the line-of-sight (LoS) and non-LoS (NLoS) components for all channels \cite{tse2005fundamentals}. For example, the channel between STAR-RIS and AP can be expressed as $\boldsymbol{G}\!=\!\sqrt{L_{\mathrm{AS}}(d)}\!\left(\!\sqrt{\frac{\kappa_{\mathrm{AS}}}{1+\kappa_{\mathrm{AS}}}} \boldsymbol{G}^{\mathrm{LoS}}\!+\!\sqrt{\frac{1}{1+\kappa_{\mathrm{AS}}}} \boldsymbol{G}^{\mathrm{NLoS}}\right)$, where $\kappa_{\mathrm{AS}}$ is the factor representing the power ratio between the LoS path and the scattered paths, $\boldsymbol{G}^{\mathrm{LoS}}$ is the LoS component modeled as the product of the steering vectors of the AP and STAR-RIS link \cite{9110912,tse2005fundamentals}, $\boldsymbol{G}^{\text {NLoS}}$ is the Rayleigh fading components with entries distributed as $\mathcal{C} \mathcal{N}(0,1)$, $L_{\mathrm{AS}}(d)$ is the distance-dependent path loss of the AP-STAR-RIS channel. We consider the following distance-dependent path loss model $L_{\mathrm{AS}}(d)=T_{0}\left(\frac{d}{d_{0}}\right)^{-\alpha_{\mathrm{AS}}}$, where $T_{0}$ is the constant path loss at the reference distance $d_{0}=1 \mathrm{~m}, d$ is the Euclidean distance between the transceivers, $\alpha_{\mathrm{AS}}$ is the path loss exponent. Since the STAR-RIS can be practically deployed in LoS with the AP, we set $\alpha_{\mathrm{AS}}=2$ and $\kappa_{\mathrm{AS}}=30 \mathrm{~dB}$ \cite{9473572,8811733}. In addition, other channels are similarly generated with $\alpha_{\mathrm{AU}}=3.67$ and $\kappa_{\mathrm{AU}}=0$ (i.e., Rayleigh fading to account for rich scattering) for the AP-user channel, $\alpha_{\mathrm{SU}}=2.5$ and $\kappa_{\mathrm{SU}}=3$ for the STAR-RIS-user channel. We consider a system with a bandwidth $1 \mathrm{MHz}$ and $T_{0}=-30 \mathrm{~dB}$. The effective noise power for the AP is $\sigma^{2}=-90~\mathrm{dBm}$. Unless specified otherwise, other parameters are set as follows: $E_{k}=10\mathrm{~J}$, $C_k=200 \mathrm{~cycles/bit}$, $\kappa_k = 10^{-25}$, and $L=1 \mathrm{~s}$.

Numerical results for the proposed method are presented in comparison with four benchmarks, including the 'Equal time allocation', ‘Conventional RIS’, the ‘Zero-forcing’ scheme with ZF beamforming for detection, and the ‘Equal energy allocation’ scheme with equally allocated energy budgets for users. In the case of 'Conventional RIS', the full-space coverage provided by the STAR-RIS is achieved by employing one conventional reflect-only RIS and one transmit-only RIS. For a fair comparison, each conventional reflect-only/transmit-only RIS is assumed to have $M/2$ elements. For equal time allocation, we divide the length of the time slot $L$ equally into two parts and let the STAR-RIS transmits the signal half the time and reflects the signal half the time.

\begin{figure}[t]
     \centering
\includegraphics[scale=0.5]{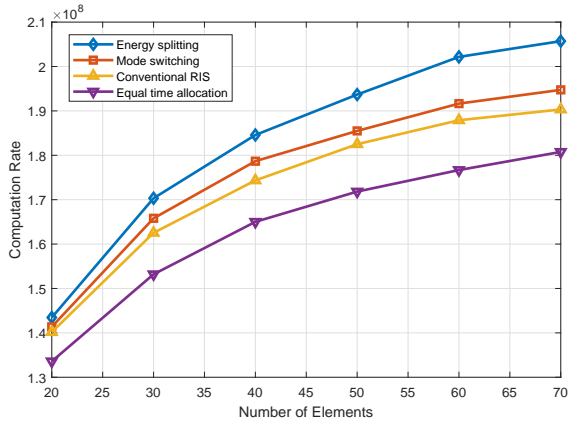}
        \caption{Computation rate versus the number of elements with $K = 8$, $N = 10$.}
        \label{Elements}
\end{figure}

In Fig. \ref{Elements}, we show the computation rate of different schemes concerning the number of STAR-RIS elements. We can observe that the computation rate of all schemes increases with the number of elements, which is consistent with the intuition that STAR-RIS with more elements has stronger channel rectification capability. It is clear that the proposed BCD optimization solution can significantly improve the performance, validating the huge benefits of deploying STAR-RIS with the joint optimization of the STAR-RIS matrix, receive beamforming, and user energy allocation. STAR-RIS with MS mode has been shown to achieve a $5\%$ increase in computation rate over the baseline of conventional RIS. Additionally, ES mode offers a performance that is nearly $10\%$ higher than MS mode. We observed that the equal time allocation scheme has the worst performance in computation rate. This is because both users in the transmission or reflection space are served by STAR-RIS only for half the duration of the time length $L$.

\begin{figure}[t]
     \centering
  \includegraphics[scale=0.5]{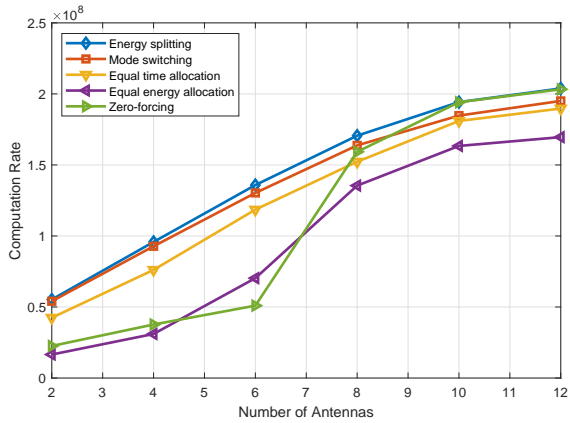}
        \caption{Computation rate versus the number of antennas at the AP $K = 8$, $M = 30$.}
        \label{Antenna}
\end{figure}

The performance of the computation rate versus the number of AP's antennas is shown in Fig. \ref{Antenna}. The STAR-RIS in ES mode has the highest computation rate, followed by the MS mode and equal time allocation scheme. As the number of antennas decreases, the performance of the zero-forcing and equal energy allocation schemes degrades dramatically. This is because zero-forcing receive beamforming fails to separate the signal stream when the number of users exceeds the number of receive antennas at the AP. In addition, since the equal energy allocation scheme cannot control the uplink transmission power, it will cause serious interference problems. Therefore, it will affect the offloading computation rate. Furthermore, we can observe that all curves of the computation rate increase as $N$ increases, and the performance gain becomes less significant as $N$ increases.
\section{Conclusion}
\label{con}
In this paper, a STAR-RIS-aided MEC system with computation offloading has been investigated. Specifically, the computation rate was maximized via the collaborative design of the STAR-RIS phase shifts, reflection and transmission amplitude coefficients, the AP's receive beamforming vectors, and the users' energy partition strategies for local computing and offloading. To solve the formulated non-convex non-continuous optimization problem, DC programming and SDR are adopted to facilitate an iterative algorithm. Numerical results have demonstrated that the STAR-RIS could significantly improve the computation rate of the system compared to the conventional RIS system.

\bibliographystyle{IEEEtran}
\bibliography{bare_conf}

\end{document}